\documentclass[preprint,12pt]{elsarticle}

\usepackage{amssymb}
\usepackage{color}

\journal{physica A}

\begin{document}

\begin{frontmatter}



\title{Self-organization and phase transition in financial markets with multiple choices}


\author{Li-Xin Zhong$^{a,b}$}\ead{zlxxwj@163.com}
\author {Wen-Juan Xu$^c$}
\author {Ping Huang$^d$}
\author {Tian Qiu$^e$}
\author {Yun-Xin He$^a$}
\author{Chen-Yang Zhong$^f$}

\address[label1]{School of Finance and Coordinated Innovation Center of Wealth Management and Quantitative Investment, Zhejiang University of Finance and Economics, Hangzhou, 310018, China}
\address[label2]{School of Economics and Management, Tsinghua University, Beijing, 100084, China}
\address[label3]{School of Law, Zhejiang University of Finance and Economics, Hangzhou, 310018, China}
\address[label4]{School of Accounting, Zhejiang University of Finance and Economics, Hangzhou, 310018, China}
\address[label5]{School of Information Engineering, Nanchang Hangkong University, Nanchang, 330063, China}
\address[label6]{Yuanpei College, Peking University, Beijing, 100871, China}

\begin{abstract}
Market confidence is essential for successful investing. By incorporating multi-market into the evolutionary minority game, we investigate the effects of investor beliefs on the evolution of collective behaviors and asset prices. It is found that the roles of market confidence are closely related to whether or not there exists another market. When there exists another investment opportunity, different market confidence may lead to the same price fluctuations and the same investment attainment. There are two feedback effects. Being overly optimistic about a particular asset makes an investor become insensitive to losses. A delayed strategy adjustment leads to a decline in wealth and one's runaway from the market. The withdrawal of the agents results in the optimization of the strategy distributions and an increase in wealth. Being overly pessimistic about a particular asset makes an investor over-sensitive to losses. One's too frequent strategy adjustment leads to a decline in wealth. The withdrawal of the agents results in the improvement of the market environment and an increase in wealth.
\end{abstract}


\end{frontmatter}


\section{Introduction}
\label{sec:introduction}

Complex phenomena are ubiquitous in the natural world and human society\cite{johnson1,zhang1,ren1,huang2,stanley1,stanley2,mantegna1}. Different from microscopic and macroscopic motions in the physical world, the price movement in financial markets is not only affected by physical factors but also psychological factors\cite{hogarth1,wiesinger1,alfi1}. In the face of a wide range of investment options, people often exhibit a strong preference for a specific investment product, which is often affected by different levels of market confidence, being optimistic, neutral or pessimistic about the market. For example, some people have a strong preference for real estate owing to collective overconfidence but not a rational analysis. Psychological studies have shown that overconfidence (underconfidence) can cause investors to underreact (overreact) to new information, especially to investment losses\cite{baker, alwathainani, soroka, moore1}. Therefore, overconfidence (underconfidence) can be defined as collective willful (denial) blindness. When  people are overconfident (underconfident) in a market, they exhibit overoptimism (overpessimism) to the market and are insensitive (oversensitive) to the losses.\cite{zhou1,podobnik1,gronlund1}.

In exploring the evolutionary dynamics of complex systems, a variety of game models have been introduced\cite{perc1,zhong4,szabo1,zheng1,galla1,challet1}. In modeling the evolutionary mechanism in financial systems, the minority game (MG) and its variant, the evolutionary minority game (EMG)\cite{johnson2,johnson3,hod1,xie,chen,quan1}, have become promising tools. In the financial markets, market impact, which measures an upward or a downward tendency of the prices subsequent to a trade buying or selling a given size, is a key factor for an agent to consider before he is ready for a shift in investment\cite{bianconi1,barato1}. Because of the sequential process of making deals, market impact may not be fully reflected in the transaction price. If one can make a trade with an immediate price, the transaction price includes no market impact. If one makes a trade with a next price, the transaction price includes full market impact.

By incorporating market impact into the MG\cite{ren2,yeung1}, Yeung et al. have investigated the effects of market impact on the evolution of stock prices. By incorporating behavioral patterns into the EMG, Zhong et al. have examined the roles of behavioral biases in the movement of stock prices\cite{zhong2,ren3}. The above studies have only told us what are the main factors affecting the stock prices in a single market, but not what roles these factors may play in multiple markets. In real financial systems, there are so many assets that an individual can invest in and how the equilibrium can be reached between different markets is a central problem. Although the multi-resource (or multi-market) problems have been studied recently \cite{huang1,wawrzyniak,ghosh1,bianconi2}, the roles of behavioral biases in the evolution of stock prices in multiple markets is still an open question. To reproduce the dynamical processes in multiple markets, in the present work, we incorporate market confidence and differently ranked markets ( assets) into the EMG. The following is our main findings.

(1) Different market confidence does not always lead to different market movement. If there are free choices for the agents to enter or withdraw from the markets, the same price fluctuations and investment attainments may occur between the markets with different levels of collective confidence.

(2) The fluctuation in asset prices is determined by the strategy distribution and the number of agents trading the asset. The clustering of the strategies leads to a large fluctuation in asset prices and the decrease in the number of agents suppresses the price fluctuation.

(3) The strategy distribution is determined by the coupled effects of strategy updating and shift in markets. For the market with collective overconfidence, the strategy distribution is mainly determined by the withdrawal of investors. For the market with collective underconfidence, the strategy distribution is mainly determined by strategy updating.

The whole paper is organized as follows. The EMG with multiple choices is introduced in section 2. In Section 3, we give simulation results and discussions about the evolution of the strategies and the asset prices. The coupled dynamics of strategy updating and shift in investment is analyzed depending upon the mean field theory in Section 4. Conclusions are summarized in Section 5.

\section{The model}
\label{sec:model}
There are total $N$ agents who repeatedly make investment in one of the two markets, market A (asset A) and market B (asset B). Originally, each agent chooses his investment asset randomly. At each time step, an agent makes his decision on buying (a=+1), selling (a=-1) or taking a holding position (a=0) according to the latest asset price information available to everyone and his investment strategy, which is denoted by a probability, called gene value, $0\leq g \leq1$. Faced with the m-bit long price information, agent i will take actions according to the prediction with probability $g$ and contrary to the prediction with probability $1-g$\cite{johnson2,johnson3}.

After all the agents have made their decisions, the stock price is determined by the following equation\cite{yeung1}

\begin{equation}
P(t+1)=P(t)+sgn[A(t)]\sqrt{\mid A(t)\mid},
\end{equation}
in which $A(t)=\sum_{i=1}^{N_A or N_B}a_i(t)$, $N=N_A+N_B$ and $a_i(t)$=+1, -1, or 0.

The market impact, $0\leq \beta \leq 1$, is incorporated into the present model in the transaction price,

\begin{equation}
P_{tr}(t)= (1-\beta)P(t)+ \beta P(t+1).
\end{equation}
For full market impact, $\beta=1$, which means that the queue is so long that the agent can only make a trade with the next price, not the immediate price he mostly appreciates,  $P_{tr}(t)= P(t+1)$. For no market impact, $\beta=0$, which means that there is no other agent before the agent and he can trade with the immediate price,   $P_{tr}(t)= P(t)$.

The attainment of agent i at a buy-sell asset transaction is determined by

\begin{equation}
B_i=P_i^{sell}-P_i^{buy},
\end{equation}
in which $P_i^{buy}$ is the transaction price he buys the asset and $P_i^{sell}$ is the transaction price he sells the asset. The accumulated attainment is defined as the wealth,

\begin{equation}
W_i=\sum_{t=1}^{t_{max}} B_i.
\end{equation}

The investment strategy evolves according to the strategy score

\begin{equation}
S=S^{+}+\gamma S^{-},
\end{equation}
in which $S^{+}=\sum B^{+}$ is the accumulated gain and $S^{-}=\sum B^{-}$ is the accumulated loss after the strategy has been adopted. There exists a threshold $S_{th}$, for $S\ge S_{th}$, one does not update his strategy. For $S< S_{th}$, one randomly chooses a new strategy from [g-$\frac{R}{2}$,g+$\frac{R}{2}$] with a reflecting boundary condition and the strategy score is reset to 0.

The parameter $\gamma$ of market confidence is incorporated into the model in the strategy score. For a highly-ranked asset, which may be the stock of a company with outstanding achievements, people exhibit collective overconfidence in the market and $0< \gamma <1$, which implies that the agents are overoptimistic about the market,  therefore, they are insensitive to the losses and underreact to new information. For an extreme case $\gamma=0$, from equation (5) we find that no matter what the losses are, $S$ is always larger than 0 and the agent will never change his strategy for $S_{th}<0$. For a lowly-ranked asset, which may be the stock of a newly listed company, people exhibit collective underconfidence in the market and $\gamma>1$, which implies that the agents are overpessimistic about the market,  therefore, they are oversensitive to the losses and overreact to new information. From equation (5) we find that $S$ is more possible to be less than 0 for a large $\gamma>1$. To understand how the market confidence affects the population behavior in the system with differently ranked markets, throughout the paper, for the agents investing in market A, the value of market confidence is $\gamma=1$, which means, people have rational market confidence in market A. By changing market confidence in market B from a small value to a large value, the evolutionary behavior in multi-market will be observed.

A shift in investment only occurs when the accumulated attainment $\omega$ is less than the threshold $\omega_{th}$.

\begin{equation}
\omega=\sum_{t=t_{in}}^{t_{out}} B,
\end{equation}
in which $t_{in}$ is the time one enters the market and $t_{out}$ is the time one  withdraws from the market. For $\omega\ge \omega_{th}$, one is possible to update his strategy but not his investment asset.

\section{Results and discussions}
\label{sec:results}
Figure 1 (a) and (b) depict the mean population $N_{B}$ investing in asset B versus market impact $\beta$ for market confidence $\gamma_{A}$=1 and different $\gamma_{B}$. For a small $\gamma_{B}$=0.1, which corresponds to the situation where people are overconfident in asset B, as $\beta$ increases from 0 to 0.5, $N_{B}$ decreases with the rise of $\beta$. As $\beta$ increases from 0.5 to 1.0, $N_{B}$ changes little with the rise of $\beta$. Increasing $\gamma_{B}$ leads to an overall increase in $N_{B}$ within the range of $0\le \beta<0.5$. For $\gamma_{B}>1$, i.e. $\gamma_{B}$=1.02, which corresponds to the situation where people are underconfident in asset B, as $\beta$ increases from 0 to 0.5, $N_{B}$ changes little with the rise of $\beta$. As $\beta$ increases from 0.5 to 1.0, $N_{B}$ decreases with the rise of $\beta$. Increasing $\gamma_{B}$ leads to an overall decrease in $N_{B}$ within the range of $0.5<\beta\le 1.0$. The transition point $\beta_{tr}\sim 0.5$ is observed.

The dependence of $N_{B}$ on $\gamma_B$ for $\beta$=0.2 and 0.8 can also be observed in fig.1 (c) and (d). For $\beta$=0.2, as $\gamma_{B}$ ranges from 0.01 to 1, $N_{B}$ increases with the rise of $\gamma_B$. A further increase in $\gamma_B$ leads to a slow decrease in $N_{B}$. For $\beta$=0.8, as $\gamma_{B}$ increases from 0.1 to 1.0, $N_{B}$ changes little with the rise of $\gamma_{B}$. As $\gamma_{B}$ increases from 1 to 8.0, $N_{B}$ firstly decreases quickly and then retains a fixed value of $N_{B}\sim 375$ with the rise of $\gamma_{B}$. The transition point $\gamma_{Btr}\sim 1.0$ is observed.

\begin{figure}
\includegraphics[width=8cm]{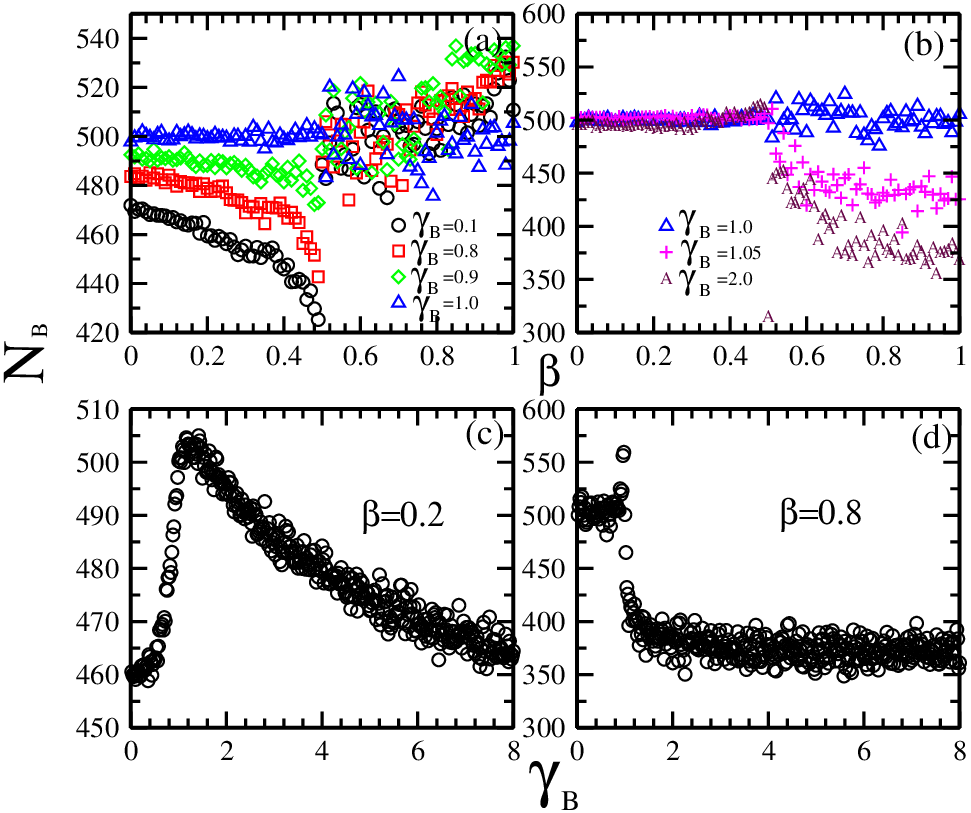}
\caption{\label{fig:epsart}The averaged number of agents $N_{B}$ investing in asset B in the final steady state as functions of $\beta$ and $\gamma_{B}$ respectively for $N=10^3$, $S_{th}= -4$, $\omega_{th}$= -200, $\gamma_{A}$=1. (a)$\gamma_{B}$=0.1 (circles), 0.8 (squares), 0.9 (diamonds), 1.0 (triangles); (b)$\gamma_{B}$=1.0 (triangles), 1.05 (pluses), 2.0 (chars);(c)$\beta$=0.2; (d) $\beta$=0.8. Final results are averaged over 100 runs and $10^3$ time steps with $10^5$ relaxation time in each run. }
\end{figure}

\begin{figure}
\includegraphics[width=12cm]{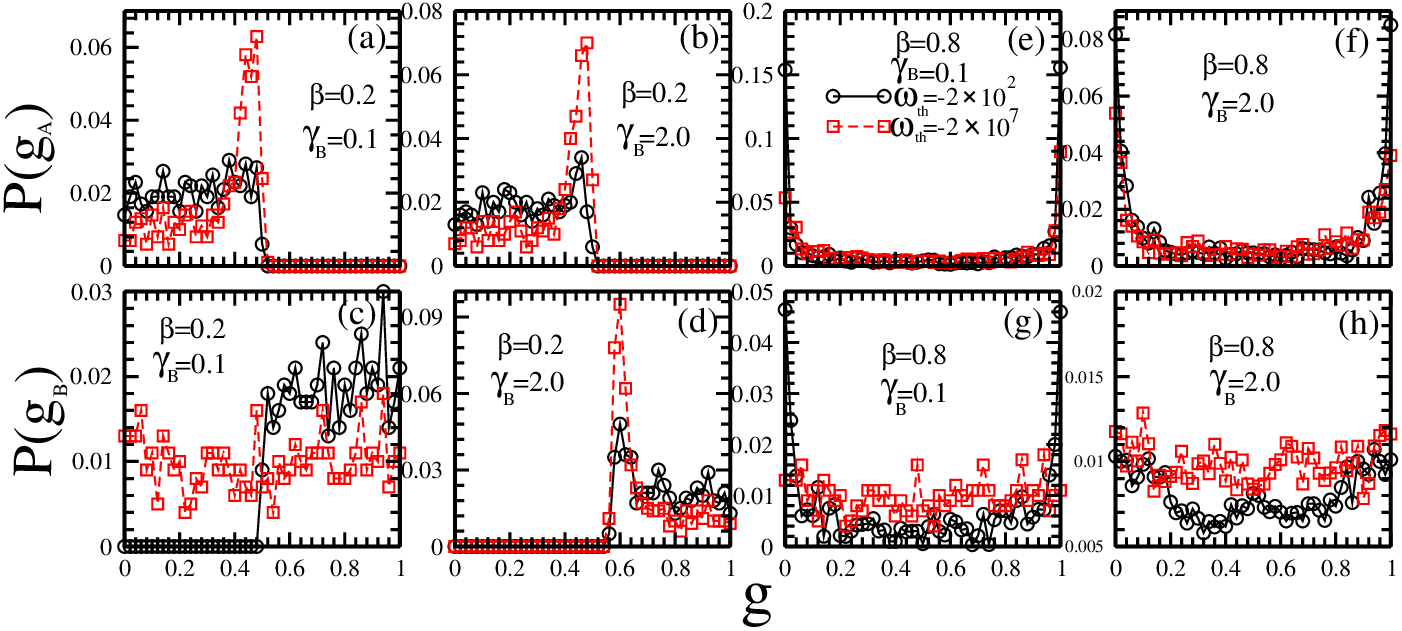}
\caption{\label{fig:epsart}The strategy distribution of the agents investing in asset A and asset B in the final steady state for $N=10^3$, $S_{th}= -4$, $\gamma_{A}$=1, $\omega_{th}=-200$ (circles), $-2\times 10^7$ (squares). (a)$P(g_{A})$ for $\beta$=0.2, $\gamma_{B}$=0.1; (b) $P(g_{A})$ for
$\beta$=0.2, $\gamma_{B}$=2.0; (c)$P(g_{A})$ for $\beta$=0.8, $\gamma_{B}$=0.1; (d) $P(g_{A})$ for $\beta$=0.8, $\gamma_{B}$=2.0; (e)$P(g_{B})$ for $\beta$=0.2, $\gamma_{B}$=0.1; (f) $P(g_{B})$ for $\beta$=0.2, $\gamma_{B}$=2.0; (g)$P(g_{B})$ for $\beta$=0.8, $\gamma_{B}$=0.1; (h) $P(g_{B})$ for $\beta$=0.8, $\gamma_{B}$=2.0. Final results are averaged over $10^3$ time steps after $10^5$ relaxation time in a run.}
\end{figure}

In fig. 2, the strategy distributions of the agents investing in asset A and asset B are plotted for $\gamma_{B}$=0.1, 2, $\beta$=0.2, 0.8  and $\omega_{th}$=-200, $-2\times 10^7$ respectively. For a small $\gamma_{B}$=0.1. When the agents can not withdraw from the markets freely, i.e.  $\omega_{th}$=$-2\times 10^7$, the strategy distribution is only determined by $\gamma_{B}$. In market A, most of the strategies cluster at $g\sim 0.49$ for $\beta$=0.2 and the strategy distribution is like a U$-$shape for $\beta$=0.8. In market B, the strategy distribution is like a uniform strategy distribution for both $\beta$=0.2 and $\beta$=0.8. When the agents can withdraw from the markets freely, i.e. $\omega_{th}=-200$, it is observed that the strategy distributions in market A and market B are nearly the same. For $\beta$=0.2, the strategies in market A (market B) are evenly distributed within the range of $g<0.5$ ($g>0.5$). For $\beta$=0.8, the strategy distributions in market A and market B are both like a U$-$shape. For a large $\gamma_{B}$=2. It is observed that the strategy distributions are nearly the same for $\omega_{th}=-200$ and $\omega_{th}=-2\times 10^7$, which implies that the strategy distributions are not affected by shift in markets.

\begin{figure}
\includegraphics[width=8cm]{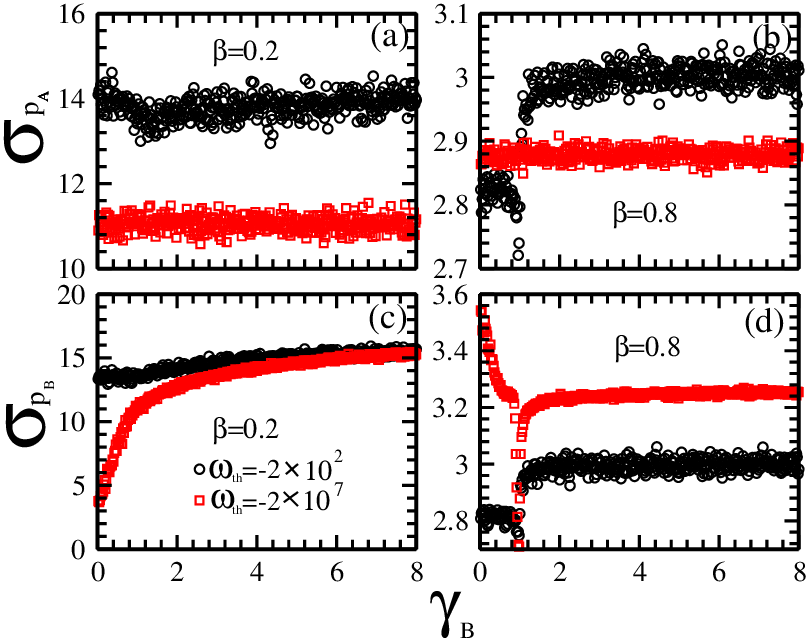}
\caption{\label{fig:epsart}The standard deviation of the prices of asset A and asset B as a function of $\gamma_B$ for $N=10^3$, $S_{th}= -4$, $\gamma_{A}$=1, $\omega_{th}=-200$ (circles), $-2\times 10^7$ (squares). (a) $\sigma_{P_A}$ for $\beta=0.2$; (b)$\sigma_{P_A}$ for $\beta=0.8$; (c) $\sigma_{P_B}$ for $\beta=0.2$; (d)$\sigma_{P_B}$ for $\beta=0.8$. Final results are averaged over 100 runs and $10^3$ time steps with $10^5$ relaxation time in each run.}
\end{figure}

Figure 3 shows the standard deviation of stock prices as a function of $\gamma_B$ for $\beta$=0.2, 0.8 and $\omega_{th}=-200$, $-2\times 10^7$ respectively. From fig. 3 (a) and (c) we observe that, for $\omega_{th}=-200$, the change of $\sigma_{P_A}$ is similar to the change of $\sigma_{P_B}$. As $\gamma_B$ increases from 0.01 to 1.0, $\sigma_{P_A}$ and $\sigma_{P_B}$ decrease with the rise of $\gamma_B$. As $\gamma_B$ increases from 1.0 to 8, $\sigma_{P_A}$ and $\sigma_{P_B}$ increase with the rise of $\gamma_B$. Comparing the results for $\omega_{th}=-200$ with the results for $\omega_{th}$=$-2\times 10^7$, we find that $\sigma_{P_A}$ ($\sigma_{P_B}$) for $\omega_{th}=-200$ is larger than that for $\omega_{th}$=$-2\times 10^7$, which implies that it should be the shift in investment that leads to the rise of $\sigma_{P_A}$ and $\sigma_{P_B}$. From fig. 3 (b) and (d) we observe that, for $\omega_{th}=-200$, the change of $\sigma_{P_A}$ is also similar to the change of $\sigma_{P_B}$. As $\gamma_B$ increases from 0.01 to 1.0, $\sigma_{P_A}$ and $\sigma_{P_B}$ change little with the rise of $\gamma_B$. As $\gamma_B$ increases from 1.0 to 8, $\sigma_{P_A}$ and $\sigma_{P_B}$ firstly have a sharp increase and then change little with the rise of $\gamma_B$. Comparing the results for $\omega_{th}=-200$ with the results for $\omega_{th}$=$-2\times 10^7$, we also find that it should be the shift in investment that leads to the similar price fluctuations in market A and market B.

\begin{figure}
\includegraphics[width=8cm]{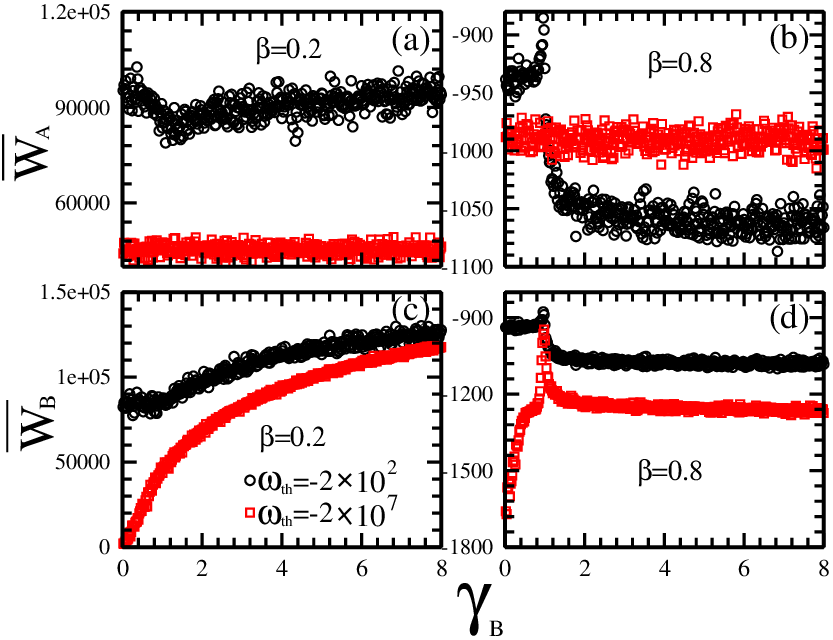}
\caption{\label{fig:epsart}The average wealth of the population investing in asset A and asset B in the final steady state as a function of $\gamma_B$ for $N=10^3$, $S_{th}=-4$, $\gamma_{A}$=1 and $\omega_{th}=-200$ (circles), $-2\times 10^7$ (squares). (a)$\overline{W_A}$ for $\beta$=0.2; (b) $\overline{W_A}$ for $\beta$=0.8; (c)$\overline{W_B}$ for $\beta=0.2$; (d) $\overline{W_B}$ for $\beta=0.8$. Final results are averaged over 100 runs and $10^3$ time steps with $10^5$ relaxation time in each run.}
\end{figure}

To understand the relationship between the attainment of the agents and the price fluctuations in market A and market B, in fig. 4 we plot the average wealth of the population investing in asset A and asset B as a function of $\gamma_B$. Comparing the results in fig. 4 with those in fig.3, we find that the average wealth is closely related to the fluctuations in asset prices. For $\beta<0.5$, a large $\sigma_{P}$ corresponds to a large $\overline{W}$. For $\beta>0.5$, a large $\sigma_{P}$ corresponds to a small $\overline{W}$.

\section{Theoretical analysis}
\label{sec:analysis}
\subsection{\label{subsec:levelA} Reasons for the existence of an equilibrium between different markets}

Consider the existence of two markets. When there is no relationship between different markets, the evolution of strategies is determined by the market confidence $\gamma$ and the market impact $\beta$. For a quite small $\gamma$, the strategy score, which satisfies $S=S^{+}+\gamma S^{-}$, will be greater than 0. Therefore, for a threshold $S_{th}<0$, the strategies will not evolve with time. A uniform strategy distribution, which is the same as the initial one, will be found within the whole range of $0\leq \gamma\leq 1$. For an intermediate or a large $\gamma$, the strategy distribution should be determined by $\beta$. For $\beta<0.5$, the strategies cluster at a particular strategy $g_0$, which increases (or decreases) with the rise of $\gamma$. For $\beta>0.5$, the strategy distribution exhibits a U$-$shape for an intermediate $\gamma$ and a uniform distribution for a large $\gamma$.

If the agents have the right to choose between different markets, the evolution of strategies will be determined by the timescale between strategy updating and shift in markets. On condition that there exists an ordinary market, market A, in which $\gamma_A=1$, and a specific market, market B, in  which $\gamma_B$ ranges from 0 to infinity. In market A, the agents using the bad strategies will update their strategies with probability $\rho_A$ and withdraw from the market with probability $\rho_A'$. In market B, the agents using the bad strategies will update their strategies with probability $\rho_B$ and withdraw from the market with probability $\rho_B'$.

Only consider a quite small $\gamma_B<1$. For $\beta<0.5$, the majority effect leads to half of the strategies are good strategies and half of the strategies are bad strategies. Suppose $\rho_A$=$\rho_A'$=0.5, $\rho_B$=0, $\rho_B'$=1. In the final steady state, $N_A$ and $N_B$ satisfy the equation

\begin{equation}
N_A-N_B=\frac{N_0(1-q^n)}{1-q},
\end{equation}
in which $N_0=\frac{N}{4}$, $q=\frac{1}{8}$ and $n\to\infty$. Therefore, for $\beta<0.5$, the strategies in the two markets are both evenly distributed within the range of $g>0.5$ (or $g<0.5$) and $N_A$ is larger than $N_B$. For $\beta>0.5$, the minority effect leads to two thirds of the strategies are good strategies and one third of the strategies are bad strategies. Suppose $\rho_A$=$\frac{2}{3}$, $\rho_A'$=$\frac{1}{3}$, $\rho_B$=0, $\rho_B'$=1. In the final steady state, $N_A$ and $N_B$ also satisfy the equation $N_A-N_B=\frac{N_0(1-q^n)}{1-q}$, in which $N_0=\frac{N}{6}$, $q=\frac{1}{27}$ and $n\to\infty$. Therefore, within the range of $\beta>0.5$, the strategies of the agents in the two markets will both self-segregate into $g\sim {0}$ and $g\sim {1}$. The difference between $N_A$ and $N_B$ for $\beta>0.5$ is less than the difference between $N_A$ and $N_B$ for $\beta<0.5$.

\subsection{\label{subsec:levelB} Coupled effects of strategy distribution and number of agents on price fluctuations}

The price return of an asset is determined by the difference between the numbers of agents buying and selling the asset,

\begin{equation}
P(t+1)-P(t)=\sqrt {\Delta N}, (\Delta N\ge 0)
\end{equation}

\begin{equation}
P(t+1)-P(t)=-\sqrt {-\Delta N}, (\Delta N< 0).
\end{equation}

The variance of the price becomes

\begin{equation}
\sigma^2=\frac{\sum_{i=t}^{t+\Delta t} \mid\Delta N_i\mid}{\Delta t}.
\end{equation}

As discussed in ref.\cite{zhong2}, $\mid\Delta N_i\mid$ is closely related to the strategy distribution. Only consider $\beta >0.5$. As the strategy distribution changes from  a uniform distribution to a U$-$shape distribution, $\mid\Delta N_i\mid$ changes from $\sqrt{N}$ to a smaller value. Therefore, when there is no relationship between different markets, the price fluctuation in market A should be lower than that in market B.

In the present model, the shift in markets is determined by the attainment in the market. After one has finished a sale, his attainment is\cite{zhong2}

\begin{equation}
P_i(t_{sell})-P_i(t_{buy})\sim f(N)(1-2\beta),
\end{equation}
in which $f(N)>0$. For $\beta >0.5$, $P_i(t_{sell})-P_i(t_{buy})<0$. A larger price fluctuation will lead to more losses. Therefore, the agents in market B are more possible to withdraw than the agents in market A. With the increase in $N_A$ and the decrease in $N_B$, the price fluctuation in market A increases and the price fluctuation in market B decreases. Finally, the price fluctuations in the two markets become the same and an equilibrium between different market is reached.

\section{Summary}
\label{sec:summary}

By incorporating market confidence into the evolutionary minority game, we have investigated the effects of overconfidence and under-confidence on the evolution of collective behaviors and asset prices. The evolution of asset prices is closely related to the strategy distribution and the number of agents in the market. The clustering of the strategies leads to a large price fluctuation and the decrease of the number of agents leads to a decline in price fluctuations. Depending upon the coupling of the changes in the strategy distribution and the number of agents, an equilibrium between different markets is reached.

In the future, the heterogeneous market confidence should be considered in the evolution of asset prices. The coupled effects of market confidence and spatial structures on the evolution of collective behaviors should be a favorite of ours.

\section*{Acknowledgments}
This work is the research fruits of National Natural Science Foundation of China (Grant Nos. 71371165, 11175079,71273224), Collegial Laboratory Project of Zhejiang Province (Grant No. Z201307),  Visiting Scholar Project of Teacher Professional Development of Zhejiang Province, Jiangxi Provincial Young Scientist Training Project (Grant No. 20133BCB23017),  Project of Department of Education of Zhejiang Province (Grant No. Y201017231), Natural Science Foundation of Zhejiang Province (Grant Nos. Y6110687, LY12G03010).





\bibliographystyle{model1-num-names}



\end{document}